\pdfoutput=1

\documentclass[10pt]{article} 
\usepackage{ragged2e}
\usepackage[utf8]{inputenc} 
\usepackage[pdftex]{graphicx}

\usepackage{geometry} 
\usepackage{graphicx} 
\usepackage{amsmath}
\usepackage{hyperref} 
\usepackage{mathtools}
\usepackage{upgreek}

\usepackage{booktabs} 
\usepackage{array} 
\usepackage{paralist} 
\usepackage{verbatim} 
\usepackage{subfig} 

\usepackage{fancyhdr} 
\usepackage{sectsty}
\allsectionsfont{\sffamily\mdseries\upshape} 

\usepackage[nottoc,notlof,notlot]{tocbibind} 
\usepackage[titles,subfigure]{tocloft} 

\title{Utility of PCA and Other Data Transformation Techniques in Exoplanet Research}
\author{Hatipoğlu, Güray}

\begin{document}
\maketitle

\centering{\section*{Abstract}}
\justifying

This paper focuses on the utility of various data transformation techniques, which might be under the principal component analysis (PCA) category, on exoplanet research. The first section introduces the methodological background of PCA and related techniques. The second section reviews the studies which utilized these techniques in the exoplanet research field and compiles the focuses in the literature under different items in the overview, with future research direction recommendations at the end. \\

\setlength{\columnsep}{3em}
\setlength{\columnseprule}{1pt}

\centering\section{Methodological Background of PCA-based Techniques}
\justifying

\centering\subsection{PCA}
\justifying

PCA is generally known as a \textit{dimensional reduction} method or a \textit{way to rotate axes to maximize variance}. Readers can look at the tutorial of Shlens (2014) and the book of Everitt and Hothorn (2011) to comprehend the theory behind PCA and its advantages/shortcomings. Below is a brief introduction to PCA: \\
\textit{Covariance matrix construction}\\
Assume that we have a dataset with N samples and M variables. Firstly, the mean of each variable will be subtracted from its corresponding sample value for every sample, otherwise, the first PC would generally contain mostly this mean value. After this, how the variables \textit{covary} with each other will be examined via a covariance matrix. The purpose behind this covariance matrix and using covariance is to see how much the initial variables vary together. The formula for this covariance is as follows:

\[
    cov_{xy} = \sigma_{xi}\sigma_{yi}= 
    \sum_{i=1}^{n} (x_{i} - \mu_{x})*(y_{i} - \mu_{y})
\]
\\

The utility is that the variables that \textit{vary together} will have higher covariance values, resulting from the multiplication of two values diverged from their mean together, compared to the cases where one variable varies but the other is near its mean. In the last case, subtracting the mean would give a near 0 value. \\
With covariances, the following covariance matrix is produced (here for four variables):

\[
\begin{bmatrix}
\sigma_{1}\sigma_{1}&\sigma_{1}\sigma_{2}&\sigma_{1}\sigma_{3} & \sigma_{1}\sigma_{4}\\
\sigma_{2}\sigma_{1}&\sigma_{2}\sigma_{2}&\sigma_{2}\sigma_{3} & \sigma_{2}\sigma_{4}\\
\sigma_{3}\sigma_{1}&\sigma_{3}\sigma_{2}&\sigma_{3}\sigma_{3} & \sigma_{3}\sigma_{4}\\
\sigma_{4}\sigma_{1}&\sigma_{4}\sigma_{2}&\sigma_{4}\sigma_{3} & \sigma_{4}\sigma_{4}
\end{bmatrix}
\]
\\
\justifying
\\

As this is a powerful way to check the overall situation between variables and many statistical methods have this in their processes, R Statistical software, MATLAB, and Python languages have a single function to create it. After this step, constructing the correlation matrix of the same dataset is easy, permitting checking the traces of multicollinearity among variables. A matrix decomposition method, eigenvalue decomposition can be applied to the covariance matrix (or correlation matrix). The entire dataset will be rotated from the original variables to a new space with \textit{principal components} ordered from the maximum explanation of variance in data to the lowest one, which is, in general, noise in the data. Every principal component is a linear combination of variables with different weights. For instance, the first PC is:\\

\[
     \alpha_{1}^{'}\textbf{x} = \alpha_{11}\textit{x}_{1} +  \alpha_{12}\textit{x}_{2} + ... +  \alpha_{1p}\textit{x}_{p} + = 
    \sum_{i=1}^{m} (\alpha_{1i}\textit{x}_{i})
\]
\\

The second PC is needed to be the one with the highest variance included in it after PC1 was extracted from the dataset, and it should be orthogonal to PC1 and all other PCs. In other words, finding the maximum variance PC after obtaining the covariance matrix (or correlation matrix) will be repeated with an added constraint of orthogonality between PCs. Further details and deriving PCs analytically have much utility to grasp the PCA methodology and better use it. It is elaborately discussed in Joliffe (2002). The common reasons to apply PCA in literature are; \\
\\
1-) reducing the number of dimensions to work on, and \\
2-) decorrelating the predictor variables so that the subsequent analyses requiring no correlation among predictors would be possible, and\\
3-) working with metavariables that the actual variables collected under them with varying coefficients. \\

Firstly, the PCA process does not inherently \textit{reduce the dimensions} of the dataset. When there are "m" variables in the dataset, PCA will generate m PCs orthogonal to each other, and as said, the first of them will have the highest variance of the dataset, and the last will probably be the random noise. There is the only rule of thumb to decide where to stop retaining PCs and discard the remaining ones, hence \textit{reduce dimensionality}. One of them is the much employed "Kaiser criterion" where only the PCs with \textit{eigen values} higher than 1 are to be kept (Kaiser, 1960). But what is eigenvalue? \\
Eigenvalue decomposition is one way to generate PCs (eigenvectors) and a set of eigenvalues that imply the total variance loaded on each PC. A square matrix (like the one we will have after generating the covariance matrix above) will be eigen-decomposed as below: \\

\centering
\[
A = Q* 
\lambda
*Q^{-1}
\]
\\
\justifying

where A above can be our covariance matrix generated above. $\lambda$ at the middle of the right-hand's side of the equation is a diagonal matrix with its diagonals containing eigenvalues, and the Q is an eigenvector matrix. Shlens (2014) and many other freely available source illustrates toy examples to make an eigendecomposition by hand and how to obtain an eigenvector matrix, or in this case, \textit{principal components}.

\centering\subsection{ICA}
\justifying

Independent Component Analysis (ICA) goes one step further than PCA (or singular value decomposition) and aims to generate \textit{statistically independent} components, in addition to the zero correlation. Shlens (2014b) provided a tutorial on independent component analysis with theory, examples, and sample code to run an efficient ICA algorithm. ICA has two steps, removing second-order correlations via SVD, then removing the remaining (assumed) correlations numerically with statistical techniques. SVD is as follows:

\centering
\[
A = U* 
\Sigma
*V^{T}
\]
\\
\justifying

U is the left singular vector matrix, V is the right singular vector matrix, and $\Sigma$ is a diagonal matrix with singular values in decreasing order from top left to right bottom. The main difference from the eigendecomposition in the PCA, outlined above, is that it can also be applied to a rectangular matrix with a different number of rows and columns. U and V matrices are orthogonal. Hence, they assist in proceeding with the decomposition of matrix A in the following way:

\centering
\[
A = U* 
\Sigma
*V^{T}, A^{T} = V* 
\Sigma^{T}
*U^{T}
\]

\justifying
\centering

\[
A^{T}A = V* 
\Sigma^{T}
*U^{T} U* 
\Sigma
*V^{T} = V* 
\Sigma^{T}\Sigma
*V^{T}
\]

\[
AA^{T} = U* 
\Sigma
*V^{T} V* 
\Sigma^{T}
*U^{T} =  U* 
\Sigma^{T}\Sigma
*U^{T}
\]

\justifying

At this point, we solve both $A^{T}A$ and $AA^{T}$ similar to the way in eigendecomposition since they are structurally similar. Moreover, the multiplication of the transpose of U and V with themselves results in identity matrix I, which is negligible in a multiplication operation. \\
According to the method in Shlens (2014b), half of the decomposition (U and $\Sigma$) is solved through a PCA-like approach for second-order correlation removal with the covariance of the data as input. Then, the remaining V matrix is numerically found with statistical techniques to ensure the singular components have the minimum dependence on each other. A point to the statistical independence:
\\

\centering
\[
I(y) = 
\int {P(y) \log_{2}{\frac{P(y)}{\prod_{i} P(y_{i})}      } dy}
\]
\justifying

This is the definition for multi-information, where the non-negative $I(y)$ value is at its lowest (0 value) if and only if all $y_{i}$ are independent of each other. After the first PCA step for decorrelation and multiplication with singular values (\textit{normalization}), we have:

 \centering
\[
\hat{s} = Vx_{w} 
\]
\justifying

where $\hat{s}$ is the source matrix, $x_{w}$ is the matrix after previous transformations, and V is the remaining rotation matrix. The critical issue here is as follows: Normally, we will return to A when U, $\Sigma$, and V are multiplied in this order. However, constructing V in such a way that maximizes the independence between the resulting components will give us different components in the inverse of the columns of the resulting combined operation ($W^{-1}$):

 \centering
\[
W = V*D^{\frac{-1}{2}}*E^{T} 
\]
\justifying

where W is the \textit{unmixing} matrix. A final note of caution is that up to now, there is no noise assumption, i.e., the components are the actual components of the data of interest, such as the voice of one person in the cocktail party and the music in the background.

\centering\subsection{SSA}
\justifying

Singular spectrum analysis (SSA) is suitable for time-series data. It does not make \textit{a priori} assumptions about the data, and it is a non-parametric method. It has many different branches, and the main form (one-dimensional) has the following steps (Hassani, 2010) \\
1-] Trajectory matrix:\\
With an \textbf{L} window length, one-dimensional time-series data becomes multi-dimensional \textit{Hankel matrix} as: \\

\[
{X} = 
\begin{bmatrix}
y_{1}&y_{2}&y_{3}& \cdots &y_{K}\\
y_{2}&y_{3}&y_{4}& \cdots&y_{K+1}\\
y_{3}&y_{4}&y_{5}& \cdots&y_{K+2}\\
\vdots &\vdots&\vdots &\ddots& \vdots\\
y_{L}&y_{L+1}&y_{L+2}&\cdots &y_{T}
\end{bmatrix}
\]

\justifying
The skew-diagonal elements are equal in this Hankel matrix. \\
2-] Singular Value Decomposition of \textit{XX}$^{T}$: \\
This will be in the form of $\textit{XX}^{T}$ = $P\Lambda P^{T}$ \\
3-] Eigen-vector Selection: \\
This is the part where we separate the one-dimensional series into different components. There are many ways to deal with this step, and one of them is grouping correlated elements into one cluster, and separating the most uncorrelated ones in different clusters (Golyandina and Korobeynikov, 2013), looking at the correlation between the reconstructed elements, (w-correlation matrix). Furthermore, if there is an \textit{a priori} knowledge about the frequency of a specific signal or noise, choosing \textit{L} window as a multiple of that frequency will assist in separating that component. More details regarding the recommendations over selecting \textit{L} window length are present in Golyandina, Nekrutkin, and Zhigljavsky (2001). \\
4-] Reconstruction of the One-dimensional time series: \\
This is via $PP^{T}$\textbf{X} after selecting components of X above in step 3.

\centering\subsection{Overview for Methodology Background}
\justifying

Since there are many variations and branching in these algorithms, it is beyond the scope of this manuscript to make an exhaustive list of possible or potential approaches. Nevertheless, PCA, SVD, ICA, and to some point SSA are basic ways to solve the problem of noise and data decomposition in the astrophysical domain. That's why summarizing these basic techniques theory before a literature review and suggesting future directions is highly beneficial. \\
The following section represents a literature review of the utilization of these techniques in the exoplanet research field, with techniques in sections, and fields of exoplanet research in subsections.

\centering\section{Literature Review}
\justifying

\centering\subsection{PCA}
\justifying

\centering\subsubsection{PCA for Circumplanetary Disk Detection and Characterization}
\justifying

Zurlo et al. (2019) aimed to benefit from the shocked and hot circumplanetary disk (CPD) mass to find proto-planets. VLT's SPHERE instrument was the data source with H$\alpha$ filter of the Zurich Imaging Polarimeter (ZIMPOL) science subsystem. H$\alpha$ narrow is 656.9 nm with 1 nm wavelength bin width, and H$\alpha$ continuum is 644.9 nm with 4.1 nm wavelength bin width. The angular spectral differential imaging (ASDI) method processed the data. They did not detect any new companion candidates, but HD99880 BaBb was resolved for the first time in the visible spectrum.

\centering\subsubsection{PCA for Detecting/Characterizing/Constraining Exoplanets}
\justifying

Davis et al. (2017) worked to amplify the strength of radial velocity resolution with a PCA-based approach in their proof-of-concept study. The idea was to disentangle the photospheric impact on absorption lines by spots and faculae on stars, and actual Keplerian Doppler shift, which will be used to characterize or constrain the characteristics of the exoplanets. Disk-integrated spectra were from Spot Oscillation and Planet code 2.0 (Dumusque et al. 2014), accounting for spots and radial velocity impacts, but not faculae. In their results, they could obtain exoplanet-originated radial velocity in a single, significant PCA, and other star-related effects with multiple different PCs. Cretignier, Dumusque, and Pepe (2022) used PCA to improve radial velocity measurements on \textit{shell time series}, which means the projection of the spectrum on the space-normalized flux versus flux gradient. They worked on HD10700 and HD128621, with successful improvements in both. One note of caution was that for HD128621, S/N should be higher than 250 to run this method smoothly. The data was High Accuracy Radial Velocity Planet Searcher (HARPS) 1D-merged spectra with data reduction software and then the YARARA pipeline preprocessed the data. 

\centering\subsubsection{PCA for Directly Imaging Exoplanets}
\justifying

Gonzales et al. (2017) constructed a vortex image processing, VIP, package in python for astronomical high-contrast imaging. It has numerous PCA-based algorithms and options, including memory-efficient incremental PCA. The main idea is retrieving PCs, and after subtracting the first components, enhancing planet or \textit{companion} signal. This is in the angular differential imaging (ADI) field, and the star should be at the center of the image for the best performance. PCA enters the scene in the point-spread function extraction post-processing step. Hunziker et al. (2017) applied PCA on high-contrast angular differential imaging (ADI) datasets to remove thermal background emission, using a clone of the PynPoint Python package. This emission is detrimental to the contrast of images from ground-based observations. They qualitatively assessed HD100546 and calculated signal to noise (S/N) ratio for HD16942 and $\beta$ Pic. This new PCA approach was more successful at datasets with strong variations in background conditions or infrequently sampled sky backgrounds. Arcidiacono and Simoncini (2018) improved ADI post-processing by replacing PCA with non-negative matrix factorization (NMF). Instead of doing PCA and removing the \textit{dazzling} star at the center and dealing with non-physical PCs, NMF only produced physically-meaningful components from the point spread function (PSF). Pogorelyuk, Kasdin, and Rowley worked on reducing residual starlight (speckle) from simulated space-based coronograph data, specifically, Fast Linearized Coronograph Optimizer (FALCO) simulation of the Wide-Field Infrared Survey Telescope (WFIRST). The method is Electric Field Order Reduction (EFOR), available at \url{https://github.com/leonidprinceton/EFOR} as a python code. The results were better than the PCA-based KLIP approach even without reference images. \\
The Speckle Calibration tool (SpeCal) for SPHERE was developed by the SPHERE consortium, which also includes PCA (Galicher et al., 2021). The way to implement PCA is based on spatial average subtraction from each wavelength separately, then projecting each frame to the N first components. The main equation to reduce stellar intensity is as follows:

\[
    \textit{R(x,y,$\theta$,$\lambda$)} = \textit{I(x,y,$\theta$,$\lambda$)} - \textit{A(x,y,$\theta$,$\lambda$)}
\]
\\
where R is reduced intensity, I is intensity before the SpeCal, and A is the speckle pattern to be removed. PCA does this removal in the following way, with spatial average subtraction:

\[
    \textit{I}_{z} \textit{(x,y,$\theta$,$\lambda$)} = \textit{I(x,y,$\theta$,$\lambda$)} - <\textit{I(x,y,$\theta$,$\lambda$)}>_{x,y}
\]

To retrieve the R in the first equation above, the averages are added back to $\textit{I}_{z}$ projected onto N first components, also known as the number of nodes. \\
Gonzales, Absil, and Droogenbroeck (2017) developed two supervised algorithms for high-contrast imaging (HCI) for better detection of specifically faint exoplanets. These were supervised exoplanet detection via direct imaging with random forest (SODIRF), and with deep neural network (SODINN). They used data from Very Large Telescope (VLT) Nasmyth Adaptive Optics System (NACO) and Spectro-Polarimetric High-contrast Exoplanet REsearch instrument (SPHERE) instruments. SODINN had a 2 to 10-factor better (depending on angular separation and dataset) true positivity rate than the classic ADI-PCA approach. They also expect far more powerful SODINN with surveys, thanks to their diverse and high-volume data. Gonzalez et al. (2016) studies low-rank plus sparse decomposition for ADI image analysis. The purpose was a higher suppression of Gaussian-assumed noise through decomposing images into low-rank, sparse, Gaussian noise components.  Their approach also targeted the drawbacks of false positive danger from speckles and removing faint companions with a non-global threshold. They used VLT/NACO in AGPM coronographic mode for the $\beta$ Pic and $\beta$ Pic b system. The approach was successful in extracting simulated companions and having better ROC curve analysis results.
Bonse, Quanz, and Amara (2021) added the temporal dimension to the post-processing of HCI, with wavelet transformation. They reported a field rotation-related improvement in S/N by up to 60 \%. The main idea behind considering wavelet or any time-domain post-processing algorithm is the fact that there is a speed difference between field rotation and variations in the speckle patterns. This difference supposedly permits filtering the noise out. Kiefer et al. (2021) employed different combinations of ADI and SDI (alone or together) with PCA to VLT-SPHERE- Integrated Field Spectrograph (IFS) instrument. The objects of interest were Beta Pictoris b, 51 Eridani b, and HR 8799 e. They found that the highest S/N was in simultaneous use of ADI and SDI (combined differential imaging, \textit{CODI}). They recommended an excess field rotation over one full-width at half maximum (FWHM) value to detect exoplanets in small separations. Samland et al. (2021) developed the Temporal Reference Analysis of Planets (TRAP) algorithm, they were able to employ more temporal dynamics-related information using all available pixels on the images in high-contrast imaging. They showcased the utility of their model from VLT-SPHERE extreme-AO fed Infra-red dual-band images (IRDIS), 51 Eridani, and $\beta$ Pictoris. SPHERE Data Center pipeline reduced the data. They significantly improved the contrast and emphasized the importance of the temporal sampling frequency. Gebhard et al. (2022) developed an algorithm based on half-sibling regression on PSF modeling and subtraction for ADI. They included more domain knowledge before the analysis not used in PCA, such as \textit{temporal order of the data, expected structure of the PSF, the causal structure of the data-generating process}.  It performed better than the PCA-based approach utilizing four datasets (Beta Pictoris L' and M', HR 8799 L', and R CrA L') from VLT/NACO. Daglayan et al. used Annular PCA on $\beta$ Pictoris dataset with not Gaussian, but Laplacian assumption of the noise owing to the exponentially decaying tail of the error term. They utilized the Python VIP package for these experiments. Their likelihood ratio mapping approach is superior to the median-signal-to-noise (SNR) and L-SNR approaches at zero false positive rate in true positive rate value.

\centering\subsubsection{PCA for Exoplanet Atmosphere}
\justifying

Ryan and Robinson (2022) applied PCA to exoplanet spectra to retrieve ocean glint from potential ocean-containing exoplanets in their crescent and other phases. The data to simulate this condition came from the Virtual Planetary Laboratory 3-D spectral earth model (Tinetti et al. 2006). Zurlo et al. (2021) simulated VLT Planet Finder SPHERE I. performance. After using the official pipeline for SPHERE, Data Reduction and Handling (DRH, Pavlov et al. 2008), 39 monochromatic 291x291 pixel size images were collected in a three-dimensional datacube. Spectral deconvolution and Karhunen-Loeve Image Projection (KLIP, Soummer et al. 2012) algorithm as a PCA-based method were implemented over this datacube. Both methods were effective in filtering out primary starlight and uncovering planets. They stated that in real cases, in addition to the hurdles coming from calibration, angular differential imaging will also be available and a library from not-planet containing light will be even better. Cruz et al. (2018) worked on constraining Qatar-1b, a close-orbiting hot Jupiter, orbital characteristics with data analysis techniques on OMEGA2000 instrument 3.5 m telescope at Calar Alto, Spain. They used PCA to remove \textit{systematic effects}, i.e. eclipse unrelated starlight. 2MASS identifier available 9 reference stars also had these \textit{systematic} effects in their PCA, and they were beneficial in this operation on Qatar-1. Damiano, Micela, and Tinetti (2019) constructed a pipeline to detect carbon monoxide (CO) and water (${H}_{2}$O) from hot Jupiter atmospheres of HD209458b and HD189733b. They modeled the exoplanetary atmosphere with \textit{T}REx code. The data source was VLT Cryogenic High-resolution Infrared Echelle Spectrography (CRIRES). Raw data processing was from the Crire kit Version-2.3.3. by European Southern Observatory (ESO), also using ESOs \textit{EsoRex} utility. They made further calibration with the SKYCALC tool to remove \textit{telluric} absorption. They applied PCA for both wavelength-domain and time-domain cases. They retained the time-domain version. ExoMol provided CO and ${H}_{2}$O lines. They applied a cross-correlation function with an injected three orders of magnitude weaker planetary-similar signal than telluric or stellar values. The cross-correlation function (CCF) they employed is as follows:
\\

\[
     CCF(d) = \frac{ \sum_{i}^{} ((x(i) - \bar{x})*(y(i-d) - \bar{y})}{\sqrt{ \sum_{i}^{} ((x(i) - \bar{x})^2}*\sqrt{ \sum_{i}^{} ((y(i-d) - \bar{y})^2}}
\]
\\
where x and y are two series, $\bar{x}$ and $\bar{y}$ are means, and d is the delay for discrete series, which CCF tries to check different time shifts for similarity, in this time-domain application. They were able to detect CO and ${H}_{2}$O in HD209458b, and only ${H}_{2}$O in the HD189733b dataset. SPEARNET network stands for \textit{Spectroscopy and Photometry of Exoplanetary Atmospheres Research Network} had a study on optimizing exoplanet atmosphere retrieval (Hayes et al., 2019). They created forward models with the PLATON package, reduced 461 wavelength bind to 10 PCA, and applied \textit{k}-means clustering over it. The clustering generated results > 99 \% reliable for James-Webb Space Telescope (JWST) and Twinkle missions and highly reliable for Hubble Space Telescope (HST). Gibson et al. (2021) further constrained the relative abundance of elements in WASP-121b high-resolution optical transmission spectroscopy. The data was from VLT-Ultraviolet-Visual Echelle Spectrograph. They tried to retrieve the full atmospheric constitution instead of cross-correlating with \textit{a priori} injected or provided components. The framework they chose for this purpose was the likelihood approach. Their atmospheric forward model was as simple as possible with also accounting for the temperature-pressure profile, and their findings were in line with the previously published results with a considerably faster framework. Matchev, Matcheva, and Roman (2022) employed unsupervised learning methods to benchmark synthetic transit spectra. They highlighted that 3D PCA or k-means clustering efficiently reduces the very-high dimensional data to three dimensions, and can account for the number of impacting gas components on its spectra. The synthetic data had ${H}_{2}$O, HCN, ${NH}_{3}$, and thick clouds. Schreier et al. (2020) utilized singular value decomposition for temperature retrieval from infrared emission spectra of synthetic planet atmospheres. Earth's atmosphere profile was from Garand et al. (2001), and earth-like G- and K-type star orbiting planet simulations of temperature and concentration profiles were from Godolt et al. (2016) with a steady-state cloud-free climate model (von Paris et al. 2015). M-dwarf star-rotation earth-like planets were from Wunderlich et al. (2019), with a 1D photochemistry model from Gebauer et al. 2018). The forward model was from P4CAtS - Python for Computational Atmospheric Spectroscopy (Schreier et al. 2019). Gaussian noise was added to the synthetic spectrum. The LW TIR (long-wave thermal infrared) was generally better, with even an S/N value of 5 that can generate less than a 10 Kelvin error value. Thatte, Deroo, and Swain (2010) developed selective principal component extraction and reconstruction for ground-based exoplanet atmosphere spectroscopy. Their method was wavelet decomposition-assisted and they used HD 189733b data from Spectropolarimeter for Planetary EXploration (SpeX) instrument at NASA Infrared Telescope Facility (IRTF). It did not require an \textit{a priori} knowledge of the planet spectrum or physical mechanisms behind the systematic errors. Their results were in agreement with Hubble Space Telescope and Spitzer data. Cabot et al. (2019) checked the robustness of the PCA-like SYSREM method to detrend the data of high-resolution exoplanet spectra and fit an air-mass function. VLT- Cryogenic High-Resolution Infrared Echelle Spectrograph (CRIRES) data of HD189733b exoplanet. They found CO, and water in 2.3 and 3.2 $\mu$m and HCN in 3.2 $\mu$m. They also cautioned the researchers about the danger of detecting false positives after choosing an incorrect location for optimizing detrending.
\\

\centering\subsection{ICA}
\justifying

\centering\subsubsection{ICA for Detecting/Characterizing/Constraining Exoplanets}
\justifying

Waldmann (2012) utilized ICA-based models to analyze a simulated signal, HD189733b and XO1b data of HST-Near Infrared Camera and Multi-Object Spectrometer (NICMOS), and Star 10118816 data of Kepler. Waldmann constructed two methods: method 1 was filtering out the signal from the entire data, and method 2 was fitting a model for the systematic noise in the data. The results of this study indicated the superior performance of the ICA-based approach in the absence of a calibration plan in the instruments or working with unknown instrument response. Morello, Waldmann, and Tinetti (2014) utilized the independent component analysis (ICA)-based technique to further constraint of the orbital and stellar parameters of the HD189733b with IR camera Infrared Array Camera (IRAC) on Spitzer Space Telescope in 3.6$\mu$m. The idea is to use wavelength-dependent transit-depth to probe the atmospheric composition of exoplanets. They reported more robust, consistent results and no evidence of stellar variability. Later, Morello (2015) studied different simulated signals to compare pixel-based ICA to polynomial centroid division and pixel-level decorrelation and found that the ICA approach is either equivalent or better than other methods with always more objectivity. Morello, Waldman, and Tinetti (2016) extended the mentioned pixel-ICA method with wavelet transformation. This modification improved the applicability of this method to the even lower S/N data, showcasing XO3b Spitzer/IRAC data analysis in the 4.5$\mu$m band. Ingalls et al. (2016) reported the repeatability, reliability, and accuracy of exoplanet eclipse depth retrieval using the post-cryogenic mission of Spitzer Space Telescope/IRAC data, 4.5 $\mu$m 10 observations of XO-3b system (both real and synthetic sets). The paper was the report of Data Challenge results, in which most participants employed the following techniques for data reduction: BiLinearly Interpolated Subpixel Sensitivity (BLISS) mapping, Kernel Regression-Decorrelation Technique (KR/Data), Gaussian Process (GP), ICA, Kernel Regression pixel mapping (KR/Pmap), Pixel Level Decorrelation (PLD), and Segmented Polynomial K2 Pipeline [SP(K2)]. Among them, BLISS, PLD, and ICA were the most accurate ones, with BLISS as the most robust method.

\centering\subsubsection{ICA for Exoplanet Atmosphere}
\justifying

Waldmann et al. (2013) applied ICA to Hubble/NICMOS data of HD189733b. It was an eclipse spectrum between 1.51 and 2.43 $\mu$m wavelengths. They report stable results after de-trending the systematic effects with the more robust ICA approach. Morello et al. (2015) studies the GJ436b atmosphere with Spitzer/IRAC transit data and the ICA-based method they previously outlined in Morello, Waldmann, and Tinetti's (2014) paper. The main challenge in this data is that the transit depth and near-Neptune size of this planet have similar actual signal characteristics to the pixel-phase signal of the instrument. ICA de-trended the time-series data of individual pixels. They did not find any stellar activity variation at $~10^{-4}$ photometric level. Damiano et al. (2018) employed ICA and also parametric methods to the near-infrared spectrum of a hot Jupiter, HAT-P-32b planet with HST/Wide Field Camera 3 (WFC3) data. \textit{T}REx code, Bayesian spectral retrieval code, generated water vapor (log ${H}_{2}$O = $-3.45^{+1.83}_{-1.65}$) and top pressure between 5.16 and 1.73 bar. ICA had lower than two factors larger magnitude of error bars. Morello (2018) reanalyzed HD 209458 Spitzer/IRAC and HST/Space Telescope Imaging Spectrograph with Stellar and Exoplanetary Atmospheres Bayesian Analysis Simultaneous Spectroscopy (SEA BASS) method. Morello had 3-5 S/N and applied pixel-ICA without wavelet preprocessing to Channels 1 and 2 of IRAC. After least-squares minimization between the reference light curve and the parametric model, Markov Chain Monte Carlo (MCMC) sampling estimated posterior distribution parameters. Channels 3 and 4 had model-fitting with the product of transit and ramp models, differently than Channels 1 and 2. HST data had ramps and after their systematic correction, MCMC-fitting took place. The approach was successful in IRAC data overall, but with HST data, there were additional complexities with the stellar limb-darkening profiles, albeit the results were still in agreement with the 3D model of Hayek et al. (2012), with deviations from 1D MARCS and 1 D ATLAS models of Knutson et al. (2007). Marcantonio et al. (2019) studied the reflected stellar signal from 51 Peg b with the ICA method. The method was successful in simulated data and real binary star systems, but low S/N in 51 Peg + 51 Peg b data did not permit a conclusive interpretation. Feinstein et al. (2022) studied WASP-39b with James Webb Space Telescope (JWST) Near Infrared Imager and Slitless Spectrograph (NIRISS) early release data. The wavelength range is from 0.6 to 2.8 $\mu$m, and the S/N is over 100. They utilized six data reduction routines and among them, \textit{iraclis} was in the PyLightcurve python package. It utilizes ExoTETHyS, SciPy, and emcee packages along the way to (1) calculate limb-darkening, (II) find the maximum likelihood model for the data, (III) remove outliers 3 standard deviation out of normalized residuals, (IV) scale uncertainties by the RMS of the normalized residuals, (V) MCMC optimization. After removing uniform contamination, the remaining non-uniform contamination around 0.72 $\mu$m was separated with ICA as another component. The results are overall satisfactory, with the main discussion on data reduction by another method provided in their paper (refer to nirHiss Python package, or "The nirHiss Pipeline" chapter in that paper).

\centering\subsection{SSA}
\justifying

\centering\subsubsection{SSA for Detecting/Characterizing/Constraining Exoplanets}
\justifying

Emilio et al. (2010) studied Be star CoRoT-ID 102761769 with \textit{cleanest} algorithm and signal spectrum analysis using CoRoT satellite in exoplanet mode. They uncovered two frequencies and assigned them as non-radial pulsations. Greco et al. (2015) suggested the singular spectrum analysis (SSA) approach to astrophysical data, including but not limited to exoplanetary detection. There were short definitions of how disentangling signal and noise can be more robust and beneficial with this method. \\
Boufleur et al. (2018) used SSA in detrending Convection Rotation and Planetary Transits (CoRoT) space mission data. A box-fitting algorithm rediscovered all known exoplanets in this dataset and suggested a dozen novel ones. SSA processed light curves without their transits, and then the eigenvalues above the noise in the lightcurve reconstructed the signal, which is a non-trivial transit-unrelated signal component. The following three CoRoT stars were under investigation with High Accuracy Radial velocity Planet Searcher (HARPS) data from the ESO Science Archive Facility: 223977153, 104848249, 652345526. Spectroscopy Made Easy (SME) analyzed reduced HARPS data to derive stellar parameters, and then, spectral synthesis of the star with the Exofast algorithm.

\centering\subsubsection{Overview for Literature Review}
\justifying

A quick wrap-up of this literature review revealed the following points:\\
1-] The most-used technique among PCA-like methods is PCA itself, also known as Karhunen-Loeve transformation in the respective studies. \\
2-] PCA has widespread application in ADI, SDI, ASDI methods, or their derivatives with VLT data for direct imaging of hot Jupiters. \\
3-] Another most applied field for PCA is the atmospheric composition retrieval in transiting planets with a library of chemical spectra and canonical correlation function assistance, and sometimes Bayesian techniques. \\
4-] ICA application is mostly confined to one research group (Waldmann, Tinetti, Morello, etc.) with Hubble Space Telescope data. \\
5-] SSA is rare, and only in some studies on CoRoT data. \\\\

Other than Bonse, Quanz, and Amara (2021), the temporal dimension was not mentioned in the direct imaging, to the best of the author's knowledge, in which several types of SSA might be beneficial. Moreover, this area is lacking in the ICA application as well. Even though planet-reflected light is not completely independent from starlight in the actual sense, the different planet-related mechanisms might alter the spectrum differently and may result in the extraction of such activities as separate components. \\
The SSA technique might be more powerful in separating signals with different frequencies, which might yield transits better.

\centering\section{References}
\justifying

A. B. Davis, J. Cisewski, X. Dumusque, D. A. Fischer, E. B. Ford. Insights on the Spectral Signatures of Stellar Activity and Planets from PCA. (2017). arXiv: 1708.00491 \\\\
A. D. Feinstein, M. Radica, L. Welbanks, C. A. Murray, K. Ohno, L-P. Coulombe, N. Espinoza, J. L. Bean, J. K. Teske, B. Benneke, et al. Early Release Science of the exoplanet WASP-39b with JWST NIRISS. (2022). arXiv: 2211.10493 \\\\
A. Pavlov, O. Möller-Nilsson, M. Feldt, et al. \textit{Society of Photo-Optical Instrumentation Engineers Conference Series}. \textit{7019}. (2008). \\\\
A. Thatte, P. Deroo, M. R. Swain. Selective principal component extraction and reconstruction: a novel method for ground-based exoplanet spectroscopy. \textit{Astronomy \& Astrophysics} \textit{523} \textbf{A35}. (2010). \\\\
A. Zurlo, A. Vigan, D. Mesa, R. Gratton, C. Moutou. et al. Performance of the VLT Planet Finder SPHERE I. (2021). arXiv: 1410.1754 \\\\
A. Zurlo, G. Cugno, M. Montesinos, S. Perez, H. Canovas, S. Casassus, V. Christiaens, L. Cieza, N. Huelamo. The widest H$\alpha$ survey of accreting protoplanets around nearby transition disks. (2019). arXiv: 1912.04911 \\\\
B. Everitt, T. Hothorn. An Introduction to Applied Multivariate Analysis with R. \textit{Springer}. (2011)\\\\
C. Arcidiacono, V. Simoncini. Approximate nonnegative matrix factorization algorithm for the analysis of angular differential imaging data. (2018). arXiv: 1807.01208 \\\\
C. A. G. Gonzales, O. Absil, M. V. Droogenbroeck. Supervised detection of exoplanets in high-contrast imaging sequences. (2017). arXiv: 1712:02841 \\\\
C. A. G. Gonzalez, O. Absil, P. -A. Absil, M. V. Droogenbroeck, D. Mawet, J. Surdej. Low-rank plus sparse decomposition for exoplanet detection in direct-imaging ADI sequences. (2016). arXiv: 1602.08381 \\\\
C. A. G. Gonzales, O. Wertz, O. Absil, V. Christiaens, D. Defrère. VIP: Vortex Image Processing Package for High-Contrast Direct Imaging. (2017). arXiv: 1705.06184 \\\\
D. J. Ryan, T. D. Robinson. Detecting Oceans on Exoplanets with Phase-Dependent Spectral Principal Component Analysis. (2022). arXiv: 2109.11062 \\\\
F. Schreier, G. G. Sebastian, P. Hochstaffl, S. Städt. Py4CAtS-PYthon for Computational ATmospheric Spectroscopy. \textit{Atmosphere} \textit{10} \textbf{5}. p. 262. (2019). \\\\
F. Schreier, S. Städt, F. Wunderlich, M. Godolt, J. L. Grenfell. SVEEEETIES: singular vector expansion to estimate Earth-like exoplanet temperatures from infrared emission spectra. \textit{Astronomy \& Astrophysics}. \textit{633} \textbf{A156}. (2020). \\\\
F. Wunderlich, M. Godolt, J. L. Grenfell, S. Städt, A. M. S. Smith, S. Gebauer, F. Schreier, P. Hedelt, H. Rauer. Detectability of atmospheric features of Earth-like planets in the habitable zone around M dwarfs. \textit{Astronomy \& Astrophysics}. \textit{624}. \textbf{A49}. (2019). \\\\
G. Greco, D. Kondrashov, S. Kobayashi, M. Ghil, M. Branchesi, C. Guidorzi, G. Stratta, M. Ciszak, F. Marino,  A. Ortolan. Singular Spectrum Analysis for astronomical time series: constructing a parsimonious hypothesis test. (2015). arXiv: 1509.03342 \\\\
G. Morello. A blind method to detrend instrumental systematics in exoplanetary light-curves. (2015). arXiv: 1503.05309 \\\\
G. Morello, I. P. Waldmann, G. Tinetti. A new look at Spitzer primary transit observations of the exoplanet HD189733b. (2014). arXiv: 1403.2874 \\\\
G. Morello, I. P. Waldmann, G. Tinetti. Repeatability of Spitzer/IRAC exoplanetary eclipses with Independent Component Analysis. (2016). arXiv: 1601.03959 \\\\
G. Morello, I. P. Waldmann, G. Tinetti. I. D. Howarth, G. Micela, F. Allard. Revisiting Spitzer transit observations with Independent Component Analysis: new results for the GJ436 system. (2015). arXiv: 1501.05866 \\\\
G. T. Tinetti, V. S. Meadows, D. Crisp, W. Fong, E. Fishbein, M. Turnbull, J. Bibring. Detectability of Planetary Characteristics in Disk-Averaged Spectra. I: The Earth Model. \textit{Astrobiology}. \textit{6}. \textbf{1}. 34-47. (2006) \\\\
H. A. Knutson, D. Charbonneau, L. E. Allen, J. J. Fortney, E. Agol, N. B. Cowan, A. P. Showman, C. S. Cooper, S. T. Megeath. A map of the day-night contrast of the extrasolar planet HD 189733b. \textit{Nature} \textit{477}. pp. 183-186. (2007). \\\\
H. Daglayan, S. Vary, F. Cantalloube, P.-A. Absil, O. Absil. Likelihood ratio map for direct exoplanet detection. (2022). arXiv: 2210.10609 \\\\
H. F. Kaiser. The application of electronic computers to factor analysis. \textit{Educational and Psychological Measurement}. \textit{20}. 141-151. (1960).\\\\
H. Hassani. A Brief Introduction to Singular Spectrum Analysis. Retrieved from: \url{https://ssa.cf.ac.uk/ssa2010/a_brief_introduction_to_ssa.pdf} (2010). \\\\
I. P. Waldmann. Of "Cocktail Parties" and Exoplanets. \textit{The Astrophysical Journal} \textit{747} \textbf{1}. pp 12-18. (2012).doi: 10.1088/0004-0637X/747/1/12 \\\\ 
I. P. Waldmann, G. Tinetti, P. Deroo, M. D. J. Hollis, S. N. Yurchenko, J. Tennyson. Blind extraction of an exoplanetary spectrum through Independent Component Analysis. (2013). arXiv: 1301.4041 \\\\
I. T. Joliffe. Principal Component Analysis. \textit{Springer}. 2nd Edition. (2002).\\\\
J. G. Ingalls, J. E. Krick, S. J. Carey, J. R. Stauffer, P. J. Lowrance, C. J. Grillmair, D. Buzasi, D. Deming, H. Diamond-Lowe, T. M. Evans, et al., Repeatability and Accuracy of Exoplanet Eclipse Depths Measured with Post-Cryogenic Spitzer. \textit{The Astronomical Journal} \textit{152} \textbf{44}. (2016). \\\\
J. J. C. Hayes, E. Kerins, S. Awiphan, I. McDonald, J. S. Morgan, P. Chuanraksasat, S. Komonjinda, N. Sanguansak, P. Kittara (SPEARNET). Optimizing exoplanet atmosphere retrieval using unsupervised machine-learning classification. (2019). arXiv: 1909.00718 \\\\
J. Shlens. A Tutorial on Independent Component Analysis. (2014b). arXiv: 1404.2986 \\\\
J. Shlens. A Tutorial on Principal Component Analysis. (2014). arXiv:1404.1100 \\\\
K. T. Matchev, K. Matcheva, A. Roman. Unsupervised Machine Learning for Exploratory Data Analysis of Exoplanet Transmission Spectra. (2022). arXiv: 2201.02696 \\\\
L. Garand, D. S. Turner, M. Larocque, J. Bates, S. Boukabara, P. Brunel, F. Chevallier, et al., Radiance and Jacobian intercomparison of radiative transfer models applied to HIRS and AMSU channels. \textit{Journal of Geophysical Research: Atmospheres}. \textit{106} \textbf{D20}. (2001). \\\\
L. Pogorelyuk, N. J. Kasdin, C. W. Rowley. Reduced Order Estimation of the Speckle Electric Field History for Space-Based Coronographs. (2022). arXiv: 1907.01801 \\\\
M. Cretignier, X. Dumusque, F. Pepe. Stellar activity correction using PCA decomposition of shells. (2022). arXiv: 2202.05902 \\\\
M. Damiano, G. Micela, G. Tinetti. A Principal Component Analysis-based method to analyze high-resolution spectroscopic data. (2019). arXiv: 1906.11218 \\\\
M. Damiano, G. Morello, A. Tsiaras, T. Zingales, G. Tinetti. Near-IR transmission spectrum of HAT-P-32b using HST/WFC3. (2018). arXiv: 1802.10010 \\\\
M. Emilio, L. Andrade, E. Janot-Pacheco, A. Baglin, J. Gutiérrez-Soto, J.C. Suarez, B. de Batz, et al. Photometric variability of the Be star CoRoT*-ID 102761769. (2010). arXiv: 1010.5576 \\\\
M. Godolt, J. L. Grenfell, D. Kitzmann, M. Kunze, U. Langematz, A. B. C. Patzer, H. Rauer, B. Stracke. Assessing the habitability of planets with Earth-like atmospheres with 1D and 3D climate modeling. \textit{Astronomy \& Astrophysics}. \textit{592} \textbf{A36}. (2016). \\\\
M. J. Bonse, S. P. Quanz, A. Amara. Wavelet-based speckle suppression for exoplanet imaging. (2021). arXiv: 1804.05063 \\\\
M. Samland, J. Bouwman, D. W. Hogg, W. Brandner, T. Henning, M. Janson. TRAP: a temporal systematics model for improved direct detection of exoplanets at small angular separations. \textit{Astronomy \& Astrophysics}. \textit{646} \textbf{A24}. (2021). \\\\
N. Golyandina, A. Korobeynikov. Basic Singular Spectrum Analysis and Forecasting with R. (2013). arXiv: 1206.6910 \\\\
N. Golyandina, V. Nekrutkin, A. Zhigljavsky. Analysis of Time Series Structure: SSA and Related Techniques. Chapmann\&Hall/CRC. (2001). \\\\
N. P. Gibson, S. K. Nugroho, J. Lothringer, C. Maguire, D. K. Sing. Relative abundance constraints from high-resolution optical transmission spectroscopy of WASP-121b, and a fast model-filtering technique for accelerating retrievals. (2021). arXiv: 2201.04025 \\\\
P. Cruz, D. Barrado, J. Lillo-Box, M. Diaz, J. Birkby, M. López-Morales, J. J. Fortney. Detection of the secondary eclipse of Qatar-1b in the Ks band. (2018). arXiv: 1608.06263 \\\\
P. D. Marcantonio, C. Morossi, M. Franchini, H. Lehmann. Using Independent Component Analysis to detect exoplanet reflection spectrum from composite spectra of exoplanetary binary systems. (2019). arXiv: 1908.10605 \\\\
P. von Paris, F. Selsis, M. Godolt, J. L. Grenfell, H. Rauer, B. Stracke. Effect of $O_{3}$ on the atmospheric temperature structure of early Mars. \textit{Icarus}. \textit{257} pp. 406-416. (2015). \\\\
R. C. Boufleur, M. Emilio, E. Janot-PAchecco, L. Andrade, S. Ferraz-Mllo, J.-D. do Nascimento Jr., R. de La Reza. A modified CoRoT detrend algorithm and the discovery of a new planetary companion. \textit{Monthly Notices of the Royal Astronomical Society} \textit{473} pp. 710-720. (2018).\\\\
R. Galicher, A. Boccaletti, D. Mesa, P. Delorme, R. Gratton, M. Langlois, et al. Astrometric and photometric accuracies in high contrast imaging: The SPHERE speckle calibration tool (SpeCal). (2021). arXiv: 1805.04854 \\\\
R. Soummer, L. Pueyo, J. Larkin.Detection and Characterization of Exoplanets and Disks Using Projections on Karhunen-Loève Eigenimages. \textit{The Astrophysical Journal}. \textit{755}. \textbf{L28}. (2012) \\\\
S. Gebauer, J. L. Grenfell, R. Lehmann, H. Rauer. Effect of Geologically Constrained Environmental Parameters on the Atmosphere and Biosphere of Early Earth. \textit{ACS Earth Space Chem.} \textit{2} \textbf{11}. pp 1112-1136. (2018). \\\\
S. H. C. Cabot, N. Madhusudhan, G. A. Hawker, S. Gandhi. On the robustness of analysis techniques for molecular detections using high-resolution exoplanet spectroscopy. \textit{Monthly Notices of the Royal Astronomical Society} \textit{482} pp. 4422-4436. (2019). \\\\
S. Hunziker, S. P. Quanz, A. Amara, M.R. Meyer. A PCA-based approach for subtracting thermal background emission in high-contrast imaging data. (2017). arXiv: 1706.10069 \\\\
T. D. Gebhard, M. J. Bonse, S. P. Quanz, B. Schölkopf. Half-sibling regression meets exoplanet imaging: PSF modeling and subtraction using a flexible, domain knowledge-driven, causal framework. (2022). arXiv: 2204.03439 \\\\
W. Hayek, D. Sing, F. Pont, M. Asplund. Limb darkening laws for two exoplanet host stars derived from 3D stellar model atmospheres. \textit{Astronomy \& Astrophysics} \textit{539}. (2012). \\\\
X. Dumusque, I. Boisse, N. C. Santos. SOAP 2.0: A Tool to Estimate the Photometric and Radial Velocity Variations Induced by Stellar Spots and Plages. \textit{The Astrophysical Journal} \textit{796}. \textbf{2}. (2014). \\\\

\end{document}